%% file: main.tex
\documentclass[runningheads]{llncs}
\usepackage{graphicx}
\usepackage{enumitem}
\setlist{nosep, leftmargin=14pt}
\usepackage{soul}
\usepackage{cite}
\usepackage{mwe} 
\usepackage{subfig}
\usepackage{censor}
\usepackage{url}

\begin{document}

%
\title{Joint optimization of a $\beta$-VAE for ECG task-specific feature extraction}
%
%
\author{Viktor van der Valk\inst{1} \and
Douwe Atsma\inst{3} \and
Roderick Scherptong\inst{3} \and
Marius Staring \inst{2}}

\authorrunning{V.O. van der Valk et al.}
%
\institute{TECObiosciences GmbH, Landshut, Germany \\ \and Leiden University Medical Center, Department of Radiology, Leiden, The Netherlands \\
\and Leiden University Medical Center, Department of Cardiology, Leiden, The Netherlands}
%
\maketitle              
\begin{abstract}
\input{Chapters/Abstract}
\end{abstract}
\section{Introduction}
\label{sec:intro}
\input{Chapters/Introduction}


\section{Methods}
\label{sec:format}

\input{Chapters/Methods}
\section{Experiments and Results}
\label{sec:pagestyle}

\input{Chapters/Results}




\section{Discussion}
\input{Chapters/Discussion}
\label{sec:majhead}

\section*{Acknowledgments}
\label{sec:acknowledgments}
\input{Chapters/Acknowledgement}




\bibliographystyle{splncs04}
\bibliography{main}

\end{document}

%% file: Chapters/Abstract.tex

Electrocardiography is the most common method to investigate the condition of the heart through the observation of cardiac rhythm and electrical activity, for both diagnosis and monitoring purposes. Analysis of electrocardiograms (ECGs) is commonly performed through the investigation of specific patterns, which are visually recognizable by trained physicians and are known to reflect cardiac (dis)function. In this work we study the use of $\beta$-variational autoencoders (VAEs) as an explainable feature extractor, and improve on its predictive capacities by jointly optimizing signal reconstruction and cardiac function prediction. The extracted features are then used for cardiac function prediction using logistic regression. The method is trained and tested on data from 7255 patients, who were treated for acute coronary syndrome at the Leiden University Medical Center between 2010 and 2021. The results show that our method significantly improved prediction and explainability compared to a vanilla $\beta$-VAE, while still yielding similar reconstruction performance.   

%% file: Chapters/Introduction.tex
The electrocardiogram (ECG), is one of the most widely used methods to analyze cardiac morphology and function, by measuring the electrical signal from the heart with multiple electrodes. ECG data is used by clinicians for both diagnostic and monitoring purposes in various cardiac syndromes. A 12-lead ECG is routinely obtained in patients to diagnose and monitor disease development. However, for the interpretation of the ECG signal, the knowledge of an expert is required. Physicians usually analyze the ECG through the recognition of specific patterns, known to be associated with disease. This however requires substantial expertise, and potentially additional relevant information exists in a 12-lead ECG missed by human interpretation. Deep learning has already proven its usefulness in the interpretation of the ECG signal in multiple classification challenges \cite{clifford2017af, alday2020classification} and more recently also in feature discovery by means of explainable AI algorithms \cite{van2022improving, van2019interpretable, Jang2021, Kuznetsov2020}. The explainablity of AI algorithms is especially valued in medical settings, where trusting a black box AI algorithm is undesirable\cite{alday2020classification}.

VAEs and in particular $\beta$-VAEs have been used as unsupervised explainable ECG feature generators in the explainable AI algorithms mentioned above \cite{higgins2017betavae}. It was shown that a $\beta$-VAE, trained on reconstruction of the ECG signal, is able to extract features from the ECG signal that can be made more interpretable by visualization of reconstructed latent space samples with the decoder of the $\beta$-VAE \cite{van2022improving}. This is a first step towards an explainable deep learning pipeline for ECG analysis. However, the features generated by a $\beta$-VAE when only trained to minimized reconstruction loss, are likely not optimal for task specific predictions.

The aim of this paper is to explore further improvement of the latent features by improving their explainability and prediction performance. We propose to improve explainability by reducing the dimension of the latent space to a level more manageable for human assessment, while encouraging outcome specific information to be captured in the small latent space, and while maintaining ECG reconstruction performance for visual assessment. To achieve this, we jointly optimize the $\beta$-VAE with a combination of a task specific prediction loss, KL-divergence and reconstruction loss. The task chosen to optimize here is left ventricular function (LVF), one of the most important determinants of prognosis in patients with cardiac disease. Current assessment of LVF requires advanced imaging methods and interpretation by a trained professional. The ECG, on the other hand, can be obtained by a patient at home. In combination with automated analysis this would facilitate remote monitoring of LVF in patients.

%% file: Chapters/Methods.tex
\subsection{Data}
To train the models for both reconstruction and LVF prediction, two datasets were used: i) A non-labeled dataset consisting of 119.886 raw 10s 12-lead ECG signals taken at 500Hz from 7255 patients diagnosed with acute coronary syndrome between 2010 and 2021 at the Leiden University Medical Center, the Netherlands; ii) A labeled dataset of 33.610 ECGs from 2736 patients of the same cohort. This dataset was labeled by visual assessment of an echocardiogram performed within 3 days before or after the ECG. The label categories, normal, mild, moderate and severe impairment were binarized for model training. When the ECG was taken within two weeks after cardiac intervention a 1-day margin was used. If a cardiac intervention was performed between ECG and echocardiography, the case was excluded. 11.5\% of the ECGs were labeled with a moderate to severe impaired LVF. The institutional review board approved the study protocol (nWMODIV2\_2022006) and waived the obligation to obtain informed consent. 

\subsection{Data preprocessing}
\begin{figure}
    \centering
    \includegraphics[width=\linewidth]{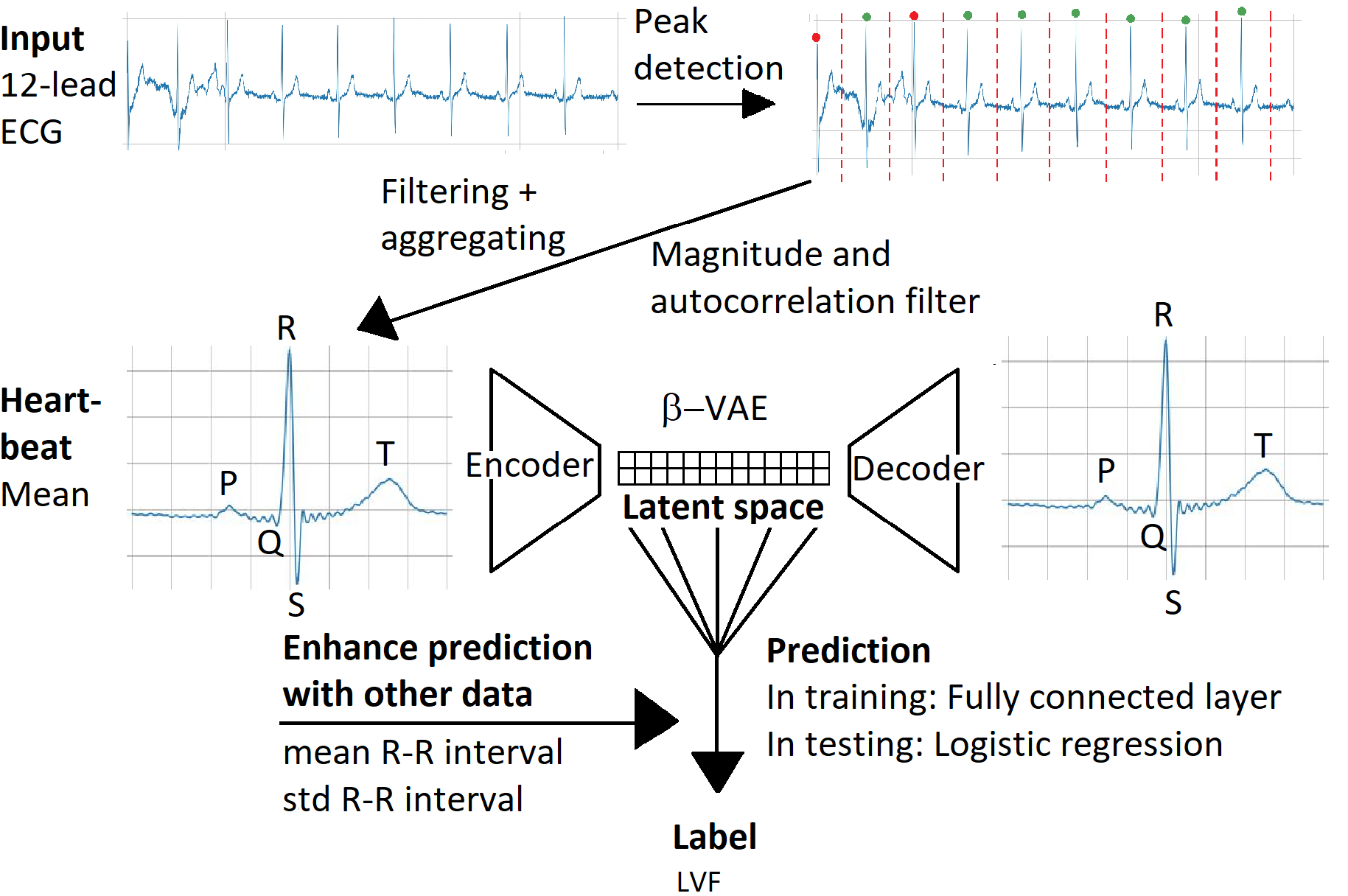}
    \caption{Preprocessing, feature extraction and prediction pipeline.}
    \label{fig:pipeline}
\end{figure}
The raw ECG signals were first split into separate heartbeats (400ms before and after the R-peak, the largest peak in the ECG, that represents depolarization of the ventricles) with a peak detection method inspired by RPNet, a U-Net structured CNN with inception blocks, that was trained on manually labeled peak locations \cite{vijayarangan2020rpnet}. The heartbeats were then filtered with a magnitude and an autocorrelation filter. The magnitude filter removed heartbeats with an average magnitude below a set threshold.
The autocorrelation filter removed signals where both the mean and maximum autocorrelation between the heartbeats were below a set threshold. These two criteria were used, since ECG signals showing multiple rhythms can result in low mean autocorrelation, but, if not noisy, will not result in low maximum autocorrelation. 
The remaining heartbeats were then averaged per ECG lead. The $\mu$ and $\sigma$ of the intervals of the subsequent R peaks were used as an additional feature for LVF prediction.

\subsection{Model overview}
To investigate a general improvement to the VAE feature extraction pipeline \cite{van2022improving, van2019interpretable, Jang2021, Kuznetsov2020}, the proposed method was tested with two architectures: i) A small VAE with 300k parameters consisting of an encoder and a mirrored decoder. Both parts contained 7 2D convolutional layers, of which 3 were residual layers, with respective channel sizes of [8,16,32,64,64,64,64] and a kernel size of 5; ii) A second larger VAE from the FactorECG pipeline as proposed by Van de Leur \textit{et al.} (2022) \cite{van2022improving} with 50M parameters. The VAEs were both extended at the bottleneck (the latent space, of size $L$), with a single fully connected layer for output prediction, in this case the LVF label, see Fig. \ref{fig:pipeline}. The $\mu$ and $\sigma$ of the RR intervals (time between two subsequent R peaks), were added to the input of the prediction layer, since the information represented by these features is lost in averaging the heartbeats. To maintain explainability of the extracted features, only one fully connected layer is used, as otherwise the features will become weighted combinations of the latent space values, which makes visualization with the decoder and subsequent interpretation complex. However, for pure prediction performance, additional fully connected layers may have been helpful. The extracted features, again with the $\mu$ and $\sigma$ of the RR interval, were subsequently analyzed with logistic regression using regularized l1 and l2 penalties on the LVF prediction task, ignoring the output of the prediction layer in the VAE. 
The VAEs were build and trained in the PyTorch 1.12 framework and trained on a Quadro RTX 6000 GPU with CUDA 11.4 \cite{NEURIPS2019_9015, cuda}, while for logistic regression we used the Scikit-learn toolbox \cite{pedregosa2011scikit}. The implementation of our models will be made publicly available via GitHub at \url{https://github.com/ViktorvdValk/Task-Specific-VAE}.


\subsection{Model training}
The $\beta$-VAE was first pretrained in a self-supervised manner with the mean heartbeats of all filtered ECG signals, minimizing i) the mean squared reconstruction error (MSE) between the input and output ECG, and ii) the KL-divergence between the output of the encoder and the standard normal distribution. The KL-divergence loss was weighted with a $\beta$ factor, like in the original paper \cite{higgins2017betavae}.
This pretrained VAE was then fine-tuned in two-steps, first the encoder and decoder were fixed, and only the prediction layer was trained, then all layers were trained end-to-end. For these fine-tuning steps, the loss function was complemented with a binary cross-entropy loss, which was weighted with a $\gamma$ factor. The \textit{task naive} VAE resulting from pretraining was compared to the \textit{task specific} VAE resulting from both fine-tuning steps.
For pretraining, both datasets were combined and split in a training (85\%) and a test set (15\%). 5-fold cross validation was done with the training set with again an 85\%:15\% ratio between training and validation set. For fine-tuning, the same procedure was used on just the labeled dataset, making sure labeled ECGs were in the same set in both cases. All data splits were grouped by patient and stratified by label in case of the labeled data splits.
Both pretraining and fine-tuning were done until convergence, i.e. until the loss on the validation set stopped improving for 25 epochs. This was done to prevent the advantage of additional training of the \textit{task specific} network.
To prevent overfitting, balanced sampling and regularization by means of drop out layers and the Adam optimizer with weight decay were used, this was especially necessary in the fine-tuning phase. To prevent gradient explosion, gradient clipping and He initialization were used \cite{he2015delving}.

\subsection{Feature evaluation}
The differences between the features from the \textit{task naive} and \textit{task specific} VAEs, were compared w.r.t.reconstruction and prediction. For reconstruction, both MSE and correlation between input and output ECG, and for prediction the Area Under the Receiver Operator Characteristic Curve (AUROC) and the macro-averaged F1 score were used. Significant difference between AUROC scores was calculated as proposed in Hanley \& McNeil (1983) \cite{hanley1983method}. 
For visualization of the representation of a latent space feature $f$ in a so called factor traversal, all features except $f$ were sampled at their mean, while $f$ was sampled in a range between $\mu - 3 \sigma$ and $\mu + 3 \sigma$. Using these samples as input for the decoder, creates a representation of that feature, which can give insight in ECG features that are important for LVF prediction.

\subsection{Baseline methods}
As a baseline method, a principal component analysis (PCA) was performed on the preprocessed ECGs, to extract features. PCA can be considered an ordered task naive linear feature extractor that focuses on the axis of the largest variance, in contrast to the VAEs which are non-ordered non-linear feature extractors, that are optimized for reconstruction.
A logistic regression predictor with just sex and age as input was used as an additional baseline.




%% file: Chapters/Results.tex
\subsection{Experiments}
The proposed pipeline contains several hyper-parameters, of which the latent space size $L$ was optimized in this study. The influence of the $\beta$ parameter was also briefly addressed. $L$ was optimized for its importance in the explainability and the reconstruction and prediction quality of the model. A higher $L$ increases the complexity of the model, and consequently decreases its explainability. An $L$ that is too low, on the other hand, restricts the capacity of the model for reconstruction and prediction. 
The PCA baseline method was considered to give an upper bound of $L$, since the number of principal components, the PCA analog for $L$, indicates how many values would be needed to capture sufficient information.

\subsection{Hyperparameter optimization}
\begin{figure}[tb]
  \centering
  \subfloat[Correlation]{\includegraphics[width=0.5\linewidth]{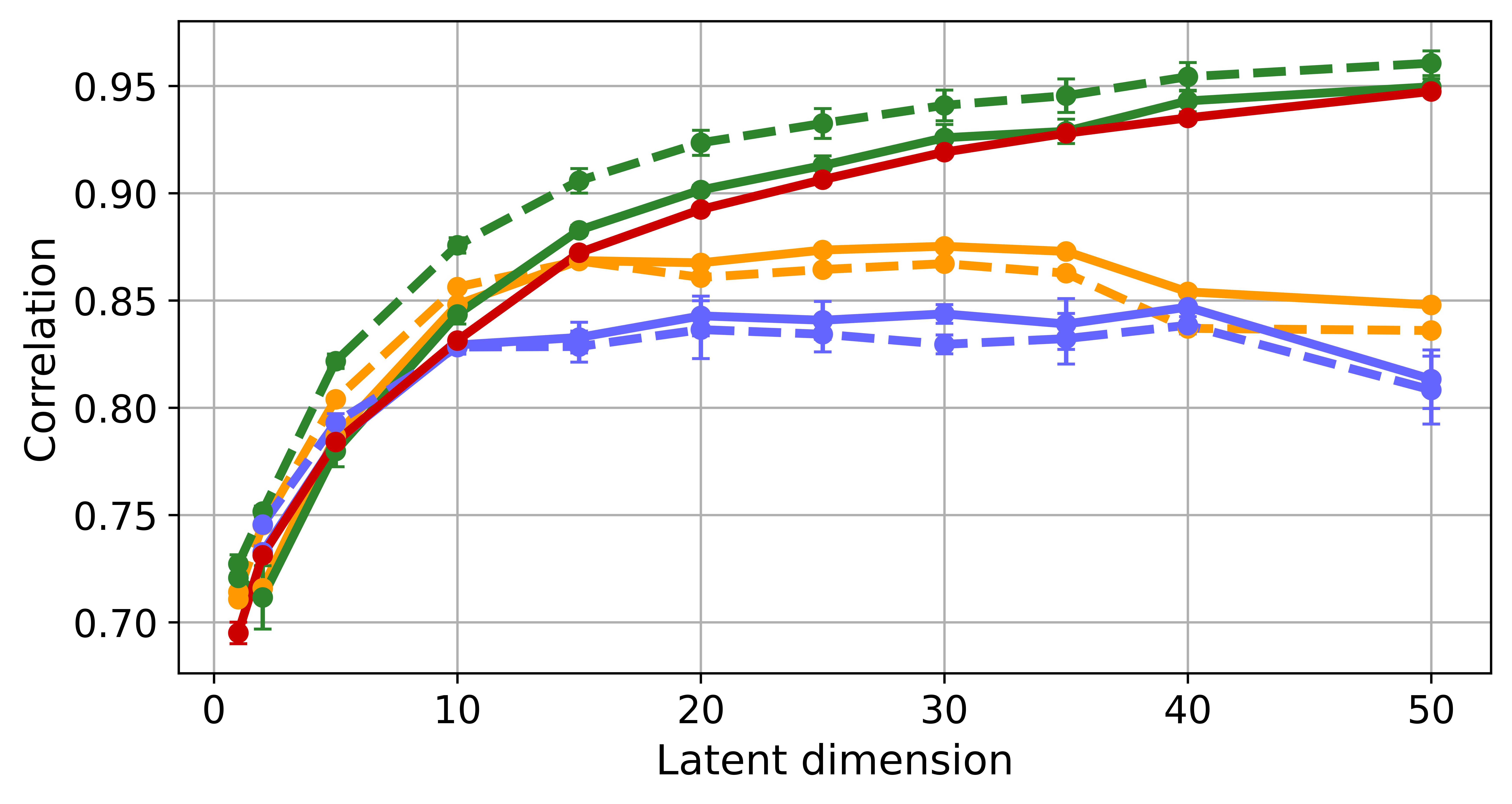}\label{fig:ld_corr}}
  \hfill
  \subfloat[MSE]{\includegraphics[width=0.5\linewidth]{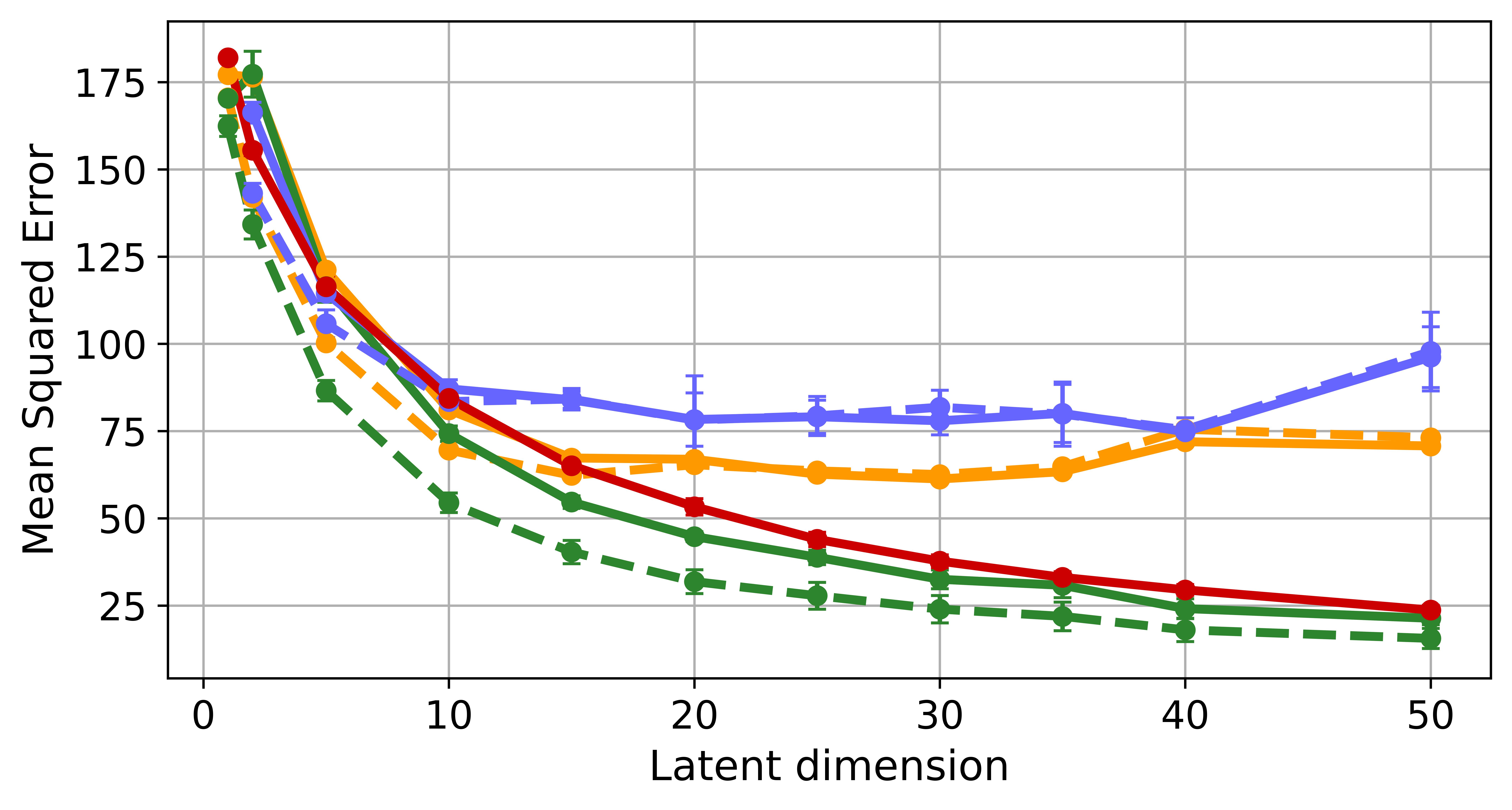}\label{fig:ld_mse}}\\

  \subfloat[AUROC]{\includegraphics[width=.5\linewidth]{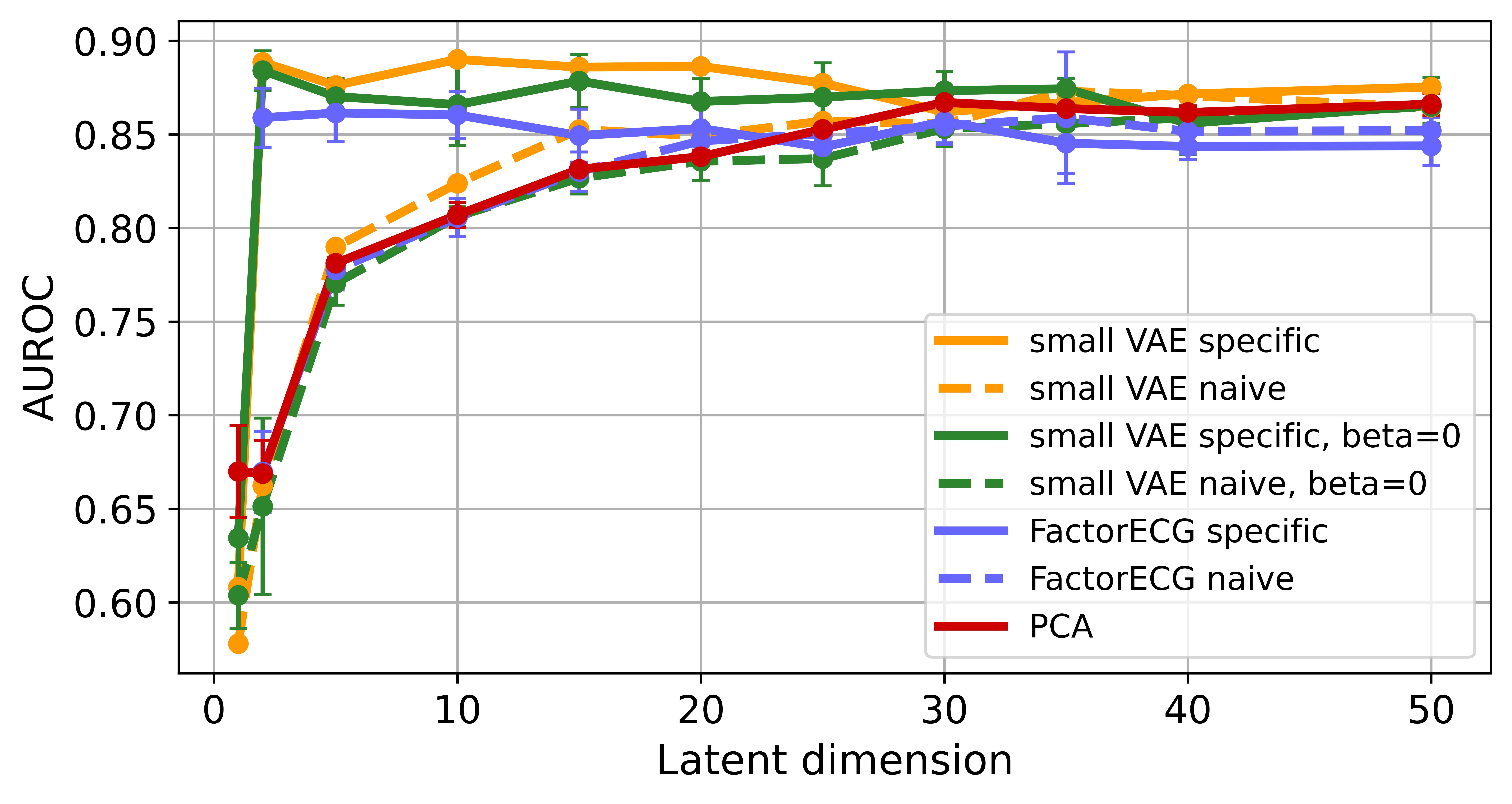}\label{fig:ld_auroc}}
  \hfill
  \subfloat[F1]{\includegraphics[width=.5\linewidth]{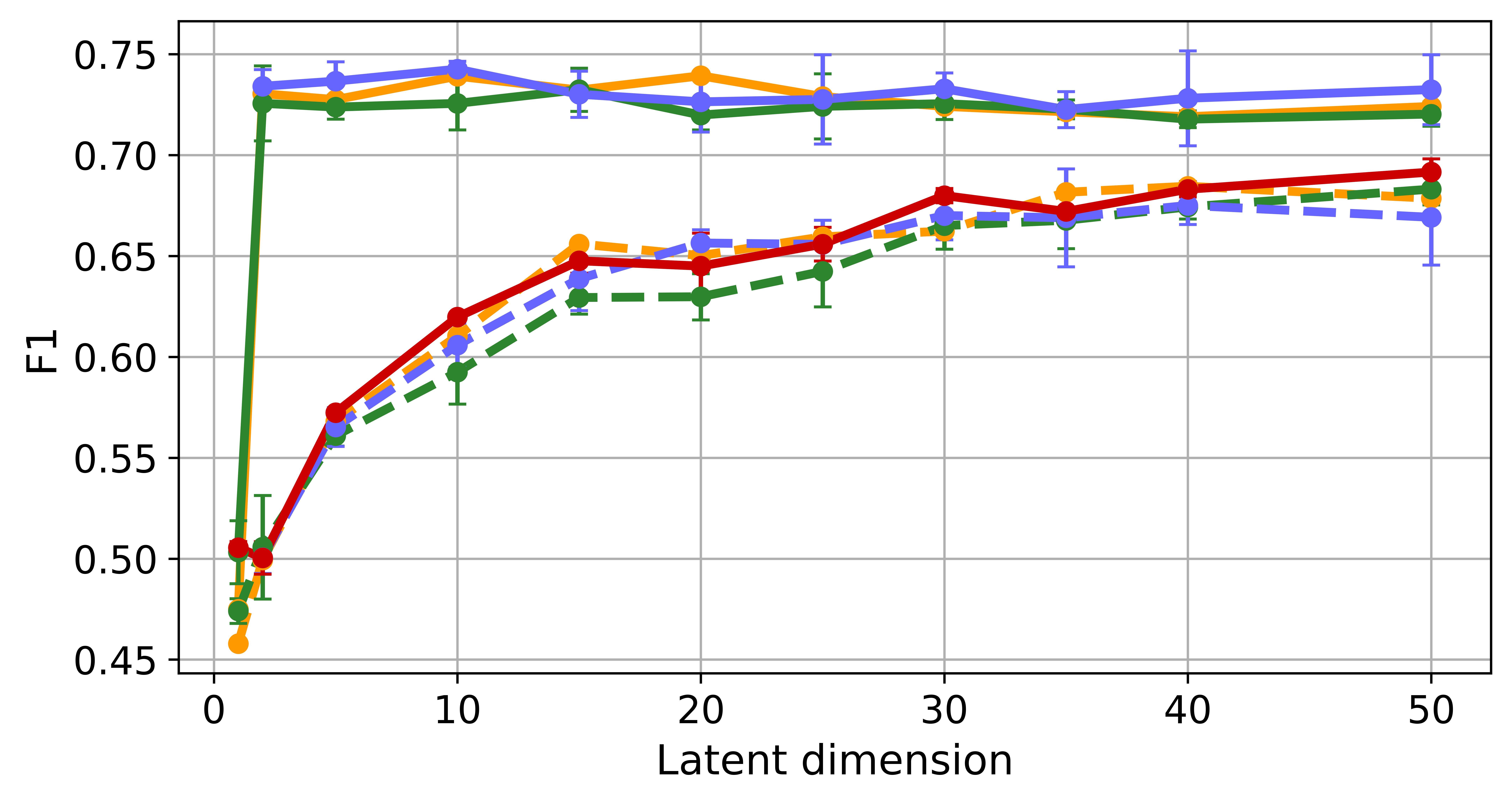}\label{fig:ld_f1}}
   \caption{Influence of the latent dimension $L$ on reconstruction quality: (a) correlation and (b) MSE, and on prediction quality: (c) AUROC and (d) F1-score, for various models. Plotted are the mean and standard deviation of 5-fold cross-validation on the validation set.}\label{fig:OptimizeL}
\end{figure}

The influences of $\gamma$ on prediction and reconstruction performance was small and was therefore fixed to 500.
The influence of $L$ on prediction quality can be seen in Fig. \ref{fig:OptimizeL}. 
The PCA baseline performs more or less equal to the \textit{task naive} networks for all $L$.
For the \textit{task specific} networks, the F1 scores are higher than their \textit{task naive} counterparts and the PCA baseline, especially for lower $L$. 
The \textit{task specific} VAEs already reached their best prediction performance starting at $L$=2, as compared to the \textit{task naive} VAEs and the PCA baseline that reach their best prediction performance from $L=30$. 
The influence of $L$ on reconstruction can be seen in Fig. \ref{fig:ld_corr} and \ref{fig:ld_mse}. All networks perform equal to the PCA baseline for low latent dimensions. The reconstruction for the small VAE and the FactorECG VAE does not seems to improve any further for respectively $L>$20 and $L>$ 15, where the PCA baseline reconstruction keeps improving with $L$. However, setting $\beta$ to 0 and thereby ablating the variational nature of the VAEs prevents this stagnation of reconstruction performance.
The \textit{task specific} networks perform equally well as their \textit{task naive} counterparts, which suggests that the additional joint optimization does not have a major negative impact on reconstruction.
The optimization shows that the relevant information for LVF prediction in the ECG signal can be captured in just two features by both VAEs. Reconstruction, on the other hand, requires at least 10/15 features for the VAEs to reach maximum performance. Therefore, in another experiment, a \textit{split task} VAE was trained, in which 8 of the latent space features where only optimized for reconstruction and only 2 also for prediction.

\subsection{Results on the test set}
\begin{figure}[tb]
  \centering
  \subfloat[\textit{task naive} with $\beta$=4]{\includegraphics[width=.33\linewidth]{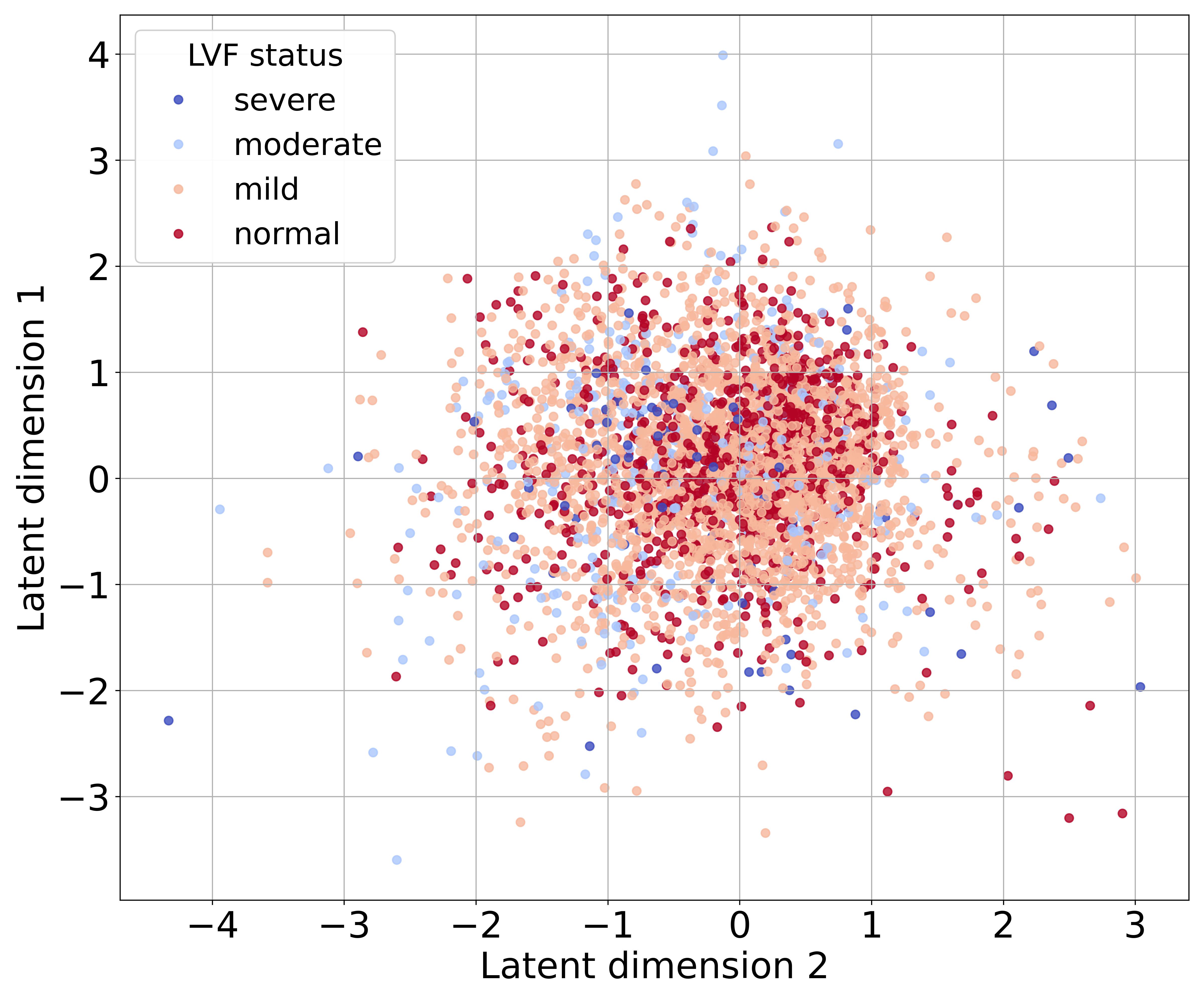}\label{fig:beta4scat_naive}}
  \subfloat[\textit{split task} $\beta$=0]{\includegraphics[width=.33\linewidth]{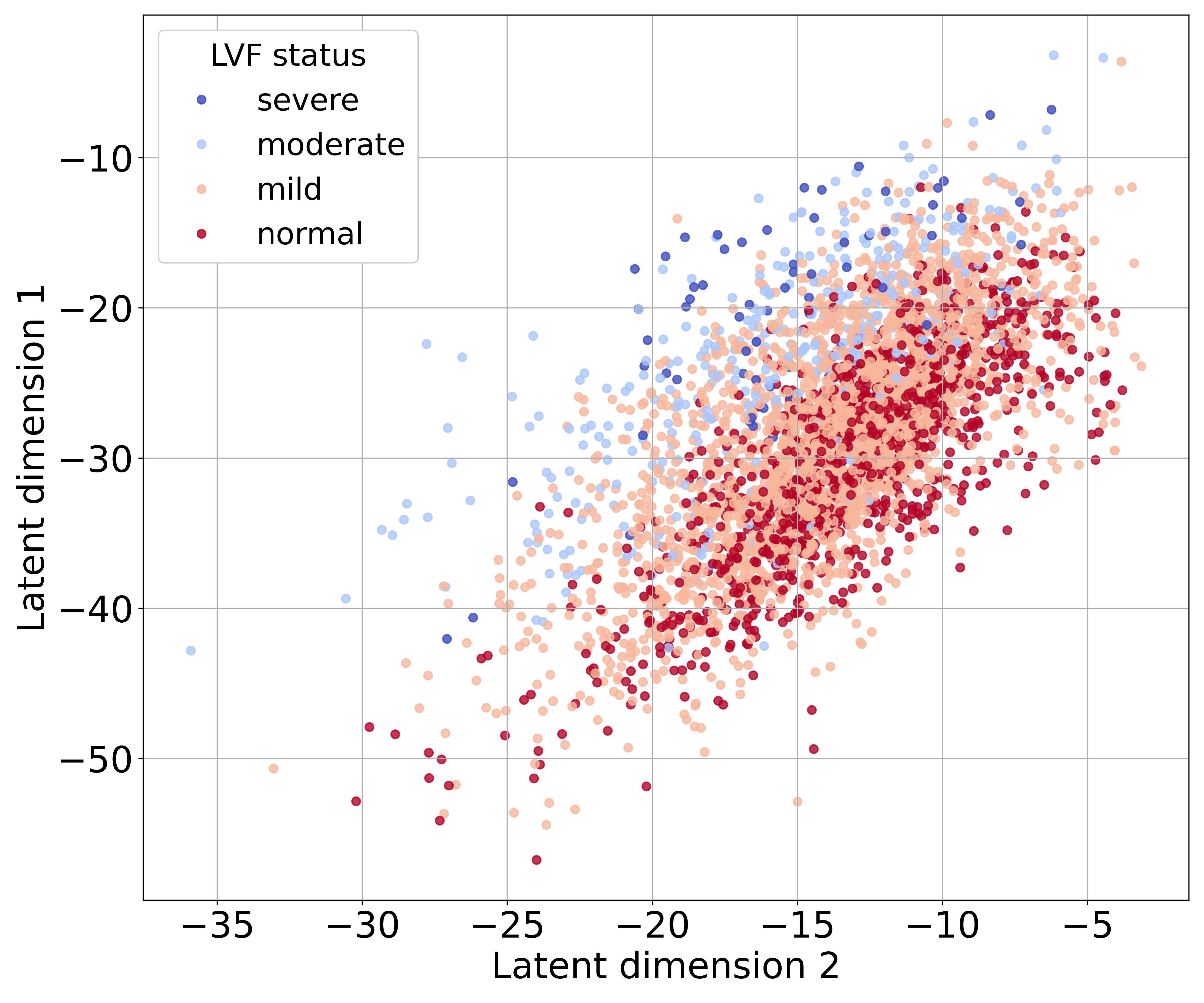}\label{fig:beta0scat}}
  \subfloat[\textit{split task} $\beta$=4]{\includegraphics[width=.33\linewidth]{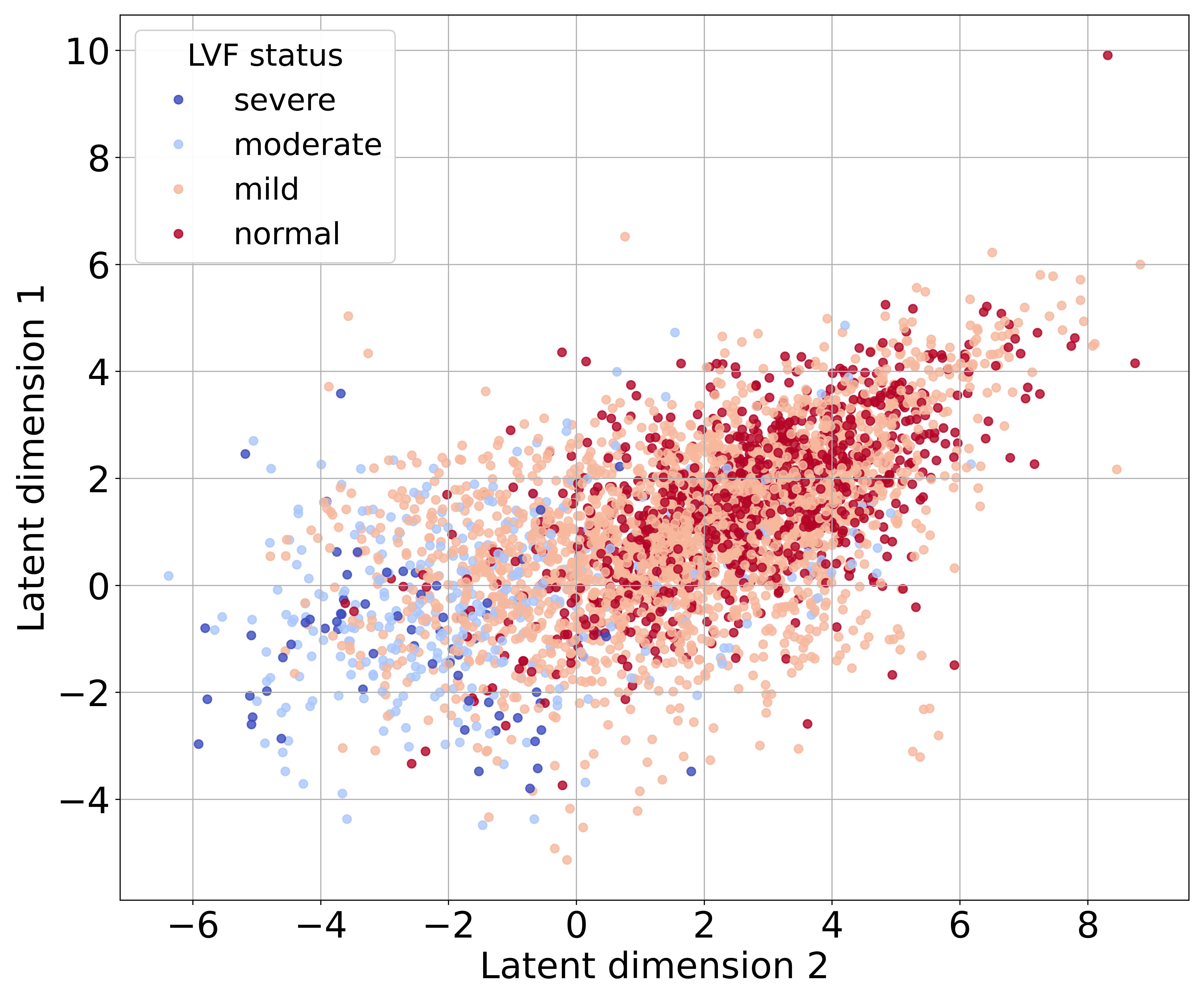}\label{fig:beta4scat}} \\
  \subfloat[Factor traversals for $\beta$=0]{
       \includegraphics[width=.24\linewidth]{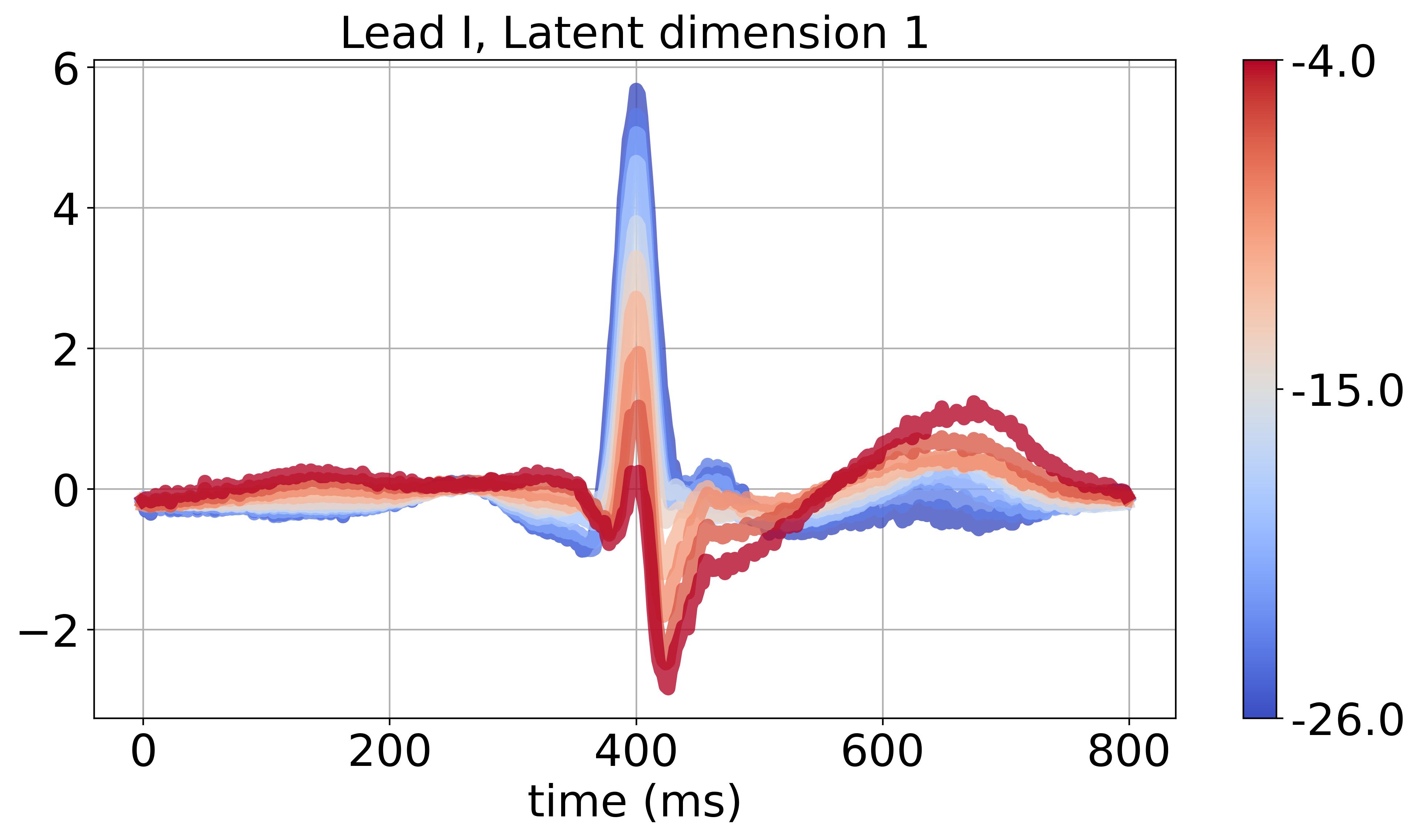}\label{fig:beta0sweep}
        \includegraphics[width=.24\linewidth]{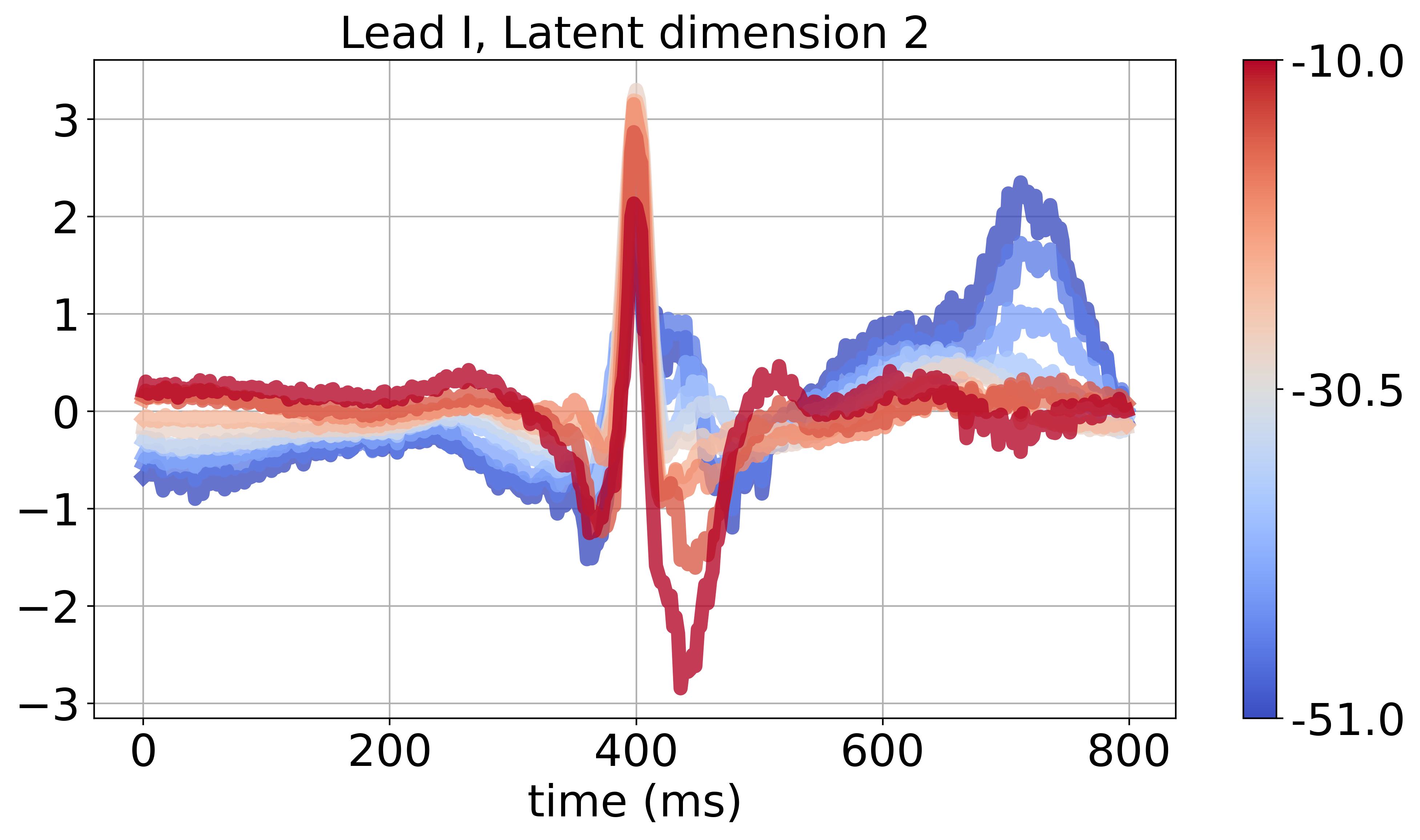}\label{fig:beta0sweep}}
  \subfloat[Factor traversals for $\beta$=4]{
        \includegraphics[width=.24\linewidth]{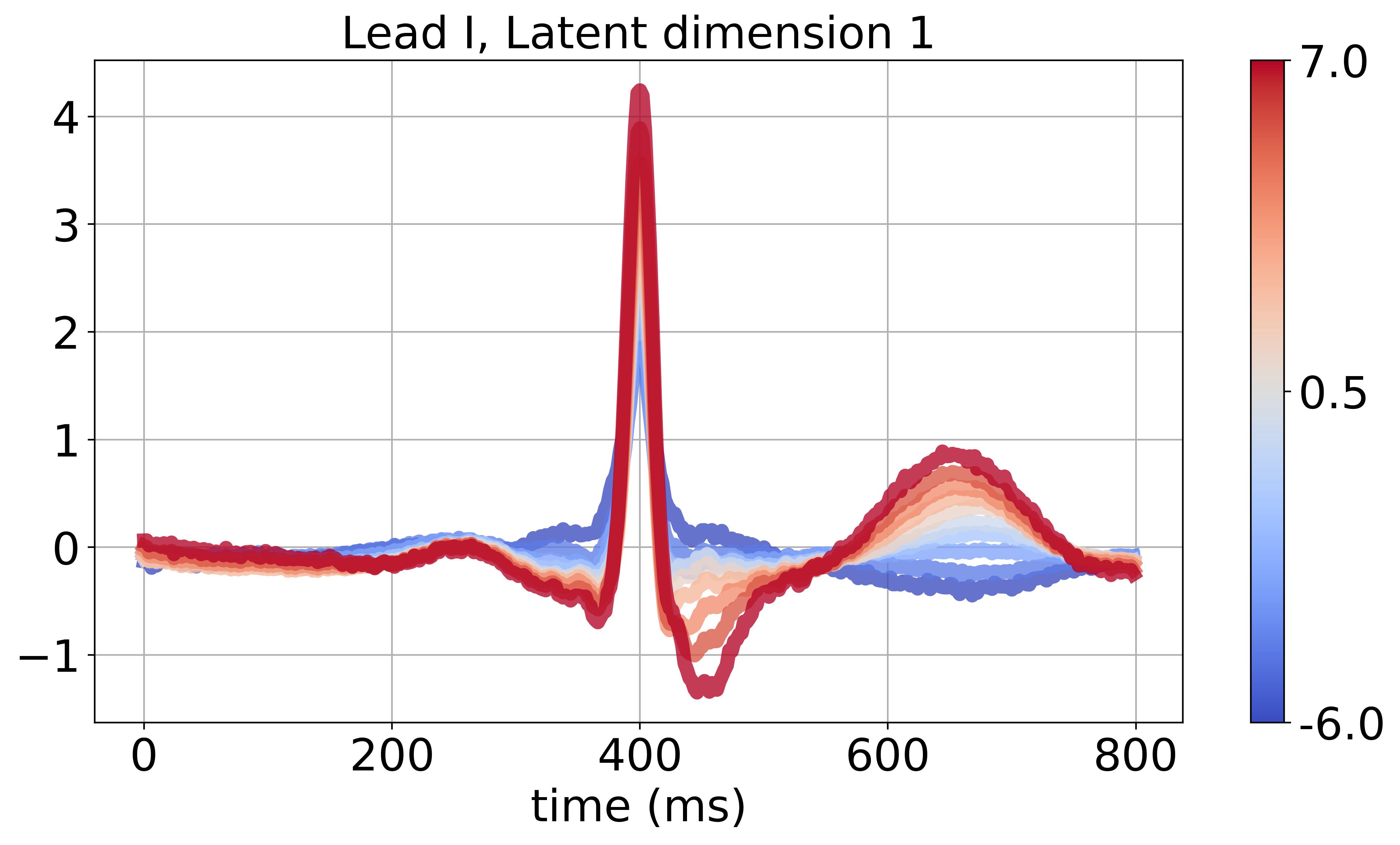}\label{fig:beta4sweep}
        \includegraphics[width=.24\linewidth]{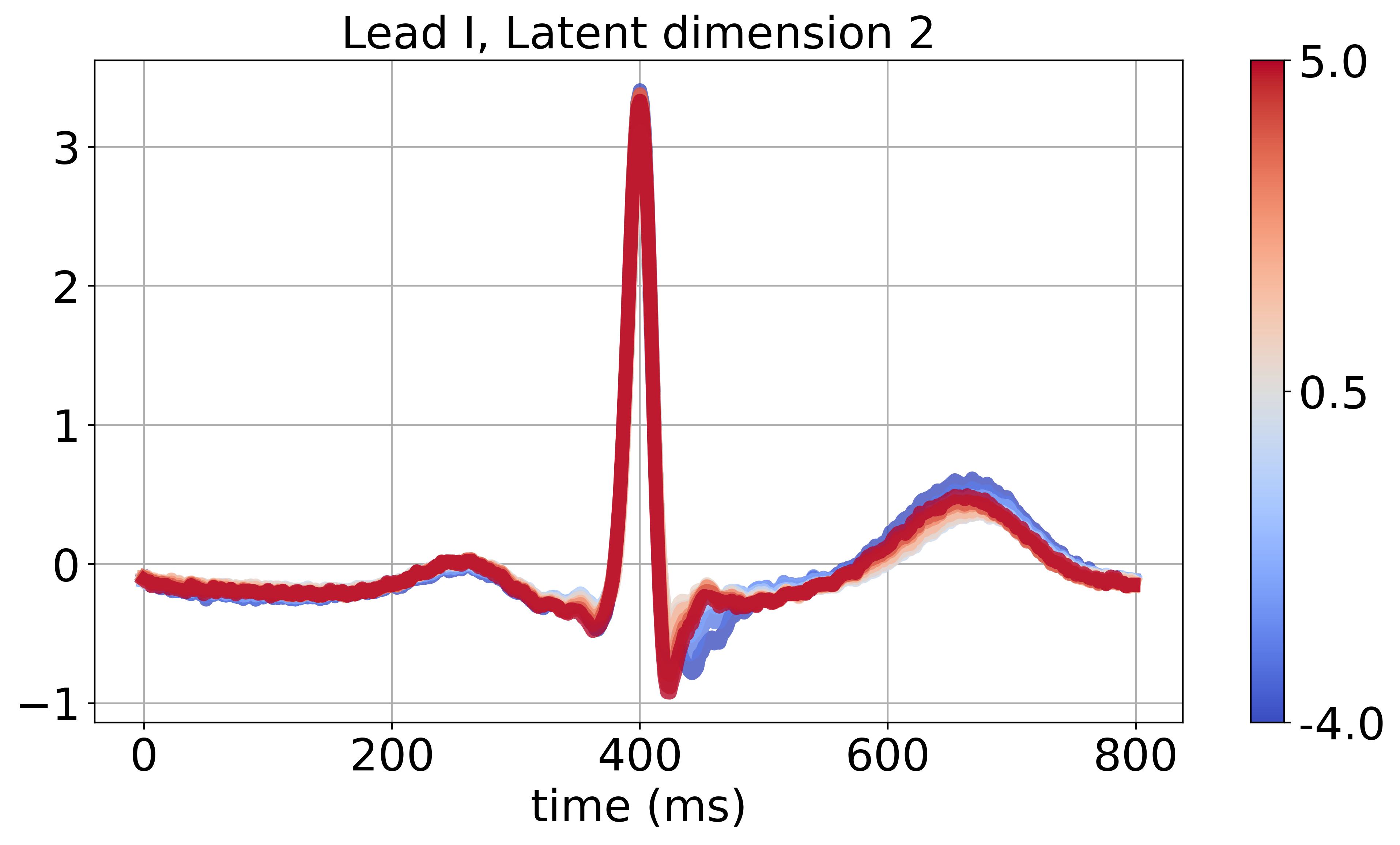}\label{fig:beta4sweep}} \\
\caption{Comparison of the latent space for different values of $\beta$, for the small VAE. For \textit{task specific} methods, the scatter plots show the two dimensions of the latent space that are optimized for prediction: (a) \textit{task naive} ($\beta$=4); (b) \textit{split task} ($\beta$=0); (c) \textit{split task} ($\beta=4$). The latent space factor traversals (d) and (e) show the visual representation of the features for Lead I of the 12-lead ECG signal: (d) $\beta$=0; (e) $\beta$=4.}\label{fig:latent_dim_vis}
\end{figure}

Table \ref{tab:comparison} shows the results on the test set for $L$=2 and $L$=10. The \textit{(split) task specific} networks significantly outperform their \textit{task naive} counterparts, the PCA baseline, and the sex and age benchmark w.r.t. LVF prediction.Fig. \ref{fig:beta4scat_naive}, \ref{fig:beta0scat} and \ref{fig:beta4scat} show that the \textit{split task}, in contrast to the \textit{task naive} small VAE with $\beta$=0 and $\beta$=4 can be used to encode the ECG signals to a landscape that visually separates the signals based on LVF status reasonably well. The factor traversals in Fig. \ref{fig:beta4sweep} and \ref{fig:beta0sweep} show an example of the interpretation of the latent features. Setting $\beta$ to 0, creates features that appear visually less informative.
\begin{table*}[tb]
    \centering
    \begin{tabular}{l||c|c|c|c|c}
    \hline
    Architecture & $L$ & MSE & Correlation & AUROC & F1 \\
    \hline
    Sex and age & 2 & - & - & 0.556 (0.520-592) & 0.474 \\
    
    \rule{0pt}{3ex}PCA & 2 & 147 & 0.724 & 0.656 (0,624-0.688) & 0.496 \\
    Small VAE task naive & 2 & 133 & 0.739 & 0.686 (0.655-0.716) & 0.503\\
    Small VAE task specific & 2 & 164 & 0.711  &  0.842* (0.822-0.861) & 0.682\\
    Small VAE split task  & 2 (10) & 76.5 &  0.838 & 0.839 (0.819-0.859) & 0.695\\  
    Small VAE split task  $\beta=0$ & 2 (10) & 73.6 &  0.838 & 0.846 (0.823 -0.862) & 0.674\\  

    FactorECG task naive & 2 & 139  & 0.735 & 0.685 (0.654-0.715) & 0.507 \\
    FactorECG task specific & 2 & 161 & 0.724 & 0.770* (0.745-0.796) & 0.695\\
    
    \rule{0pt}{3ex}PCA & 10 & 77.2 & 0.826 & 0.761 (0.735-0.787) & 0.580 \\

    Small VAE task naive & 10 & 63.1 & 0.854 & 0.803 (0.781-0.826) & 0.586\\
    Small VAE task specific & 10 & 70.6 & 0.847  &  0.853* (0.834-0.871) & 0.679\\
    
    FactorECG task naive & 10 & 84.6  & 0.820 & 0.770 (0.745-0.796) & 0.579 \\
    FactorECG task specific & 10 & 87.2 & 0.822 & 0.833* (0.813-0.854) & 0.707\\
  

    
    \hline
    \end{tabular}
    \caption{Reconstruction and LVF prediction comparison on the test set for the \textit{task naive} and \textit{task specific} architectures. The results show the $\mu$ of 5-fold cross-validation. AUROC is shown with its 95\% confidence interval.* p-value $<$ 0.01 between AUROC of task naive and specific method for all folds.}
    \label{tab:comparison}
\end{table*}


%% file: Chapters/Discussion.tex
Joint optimization of a $\beta$-VAE successfully generated features that contain more information about LVF, without hampering reconstruction of the ECG signal. We hypothesize that the $\beta$-VAEs have multiple optima for ECG reconstruction of which only some generate features that are relevant for LVF prediction. This study shows that joint optimization will favor this desired subset of optima, and that this is true for different architectures. In addition, we showed that jointly optimizing only a subset of the latent space features for prediction, results in aggregation of the predictive information, thereby improving explainability.

The AUROC score of the FactorECG VAE prediction is similar when compared to van der Leur \textit{et al.} (2022) \cite{van2022improving} (AUROC$\approx$0.9 for $L$=36). However, the proposed small VAE achieved equal if not better reconstruction and prediction performance with less than 1\% of the parameters as shown in Fig. \ref{fig:OptimizeL}.

The F1 score is considered more robust than the AUROC score with data imbalance, which is the case here \cite{jeni2013facing}. From Fig \ref{fig:ld_f1} we can therefore conclude that the \textit{task specific} networks outperform the \textit{task naive} networks for any $L$. The differences between the \textit{task specific} networks and their \textit{task naive} versions in prediction, at similar reconstruction, indicate that the ECG signal can be summarized with a set of latent features of which only a subset is important for LVF prediction. The joint optimization promotes the extraction of this subset especially when $L$ is small. Figs. \ref{fig:ld_mse} and \ref{fig:ld_corr} show that the PCA baseline outperforms both VAEs in reconstruction for $L>$ 20 when $\beta$=4, but not for $\beta$=0. This indicates that the VAEs are restricted in reconstruction by the KL-divergence loss. This loss was shown to promote feature disentanglement and a gradient in the latent space \cite{mathieu2019disentangling}. Figs \ref{fig:beta0sweep} and \ref{fig:beta4sweep} show that without this loss ($\beta$=0) the latent features are more complex to interpret. This could be explained as a reduction of the disentanglement of the features resulting from the absence of the KL-divergence loss. However, Figs \ref{fig:beta4scat} and \ref{fig:beta0scat} both show a gradient in the latent space, which suggests that the prediction loss on its own also promotes a gradient in the latent space. 
Moreover, Fig \ref{fig:beta4scat} shows dependence, and thus a lack of disentanglement, between the latent features even when $\beta$=4. This complex interplay between the three losses used in the joint optimization, is very relevant for the explainability aspect of this method, but beyond the scope of the current study. We aim to examine the complex interplay in future work.
In conclusion, the proposed joint optimization improves both explainability and prediction performance of VAEs by extraction of a smaller set of LVF specific features from the ECG signal. This could reduce the need of more advanced imaging methods, currently needed to measure the LVF. This opens the way for remote monitoring of left ventricular function in patients.   


%% file: Chapters/Acknowledgement.tex
This project has received funding from the European Union's Horizon 2020 research and innovation programme under the Marie Sklodowska-Curie grant agreement No 860173.